Leveraging High-Activity Post-Therapeutic I-131 Scans for Deep Learning Reconstruction of Extremely Sparse Planar Images in Differentiated Thyroid Cancer

Boyang Pan[1,2], Hang Yang[3], Jianchen Pan[3], Cui Gan[4], Lanxi Zhang[1,2], Hongmin Li[3], Nan-Jie Gong[1,2]
[1]Institute of Magnetic Resonance and Molecular Imaging in Medicine, East China Normal University, Shanghai, China
[2]Shanghai Key Laboratory of Magnetic Resonance, School of Physics and Electronic Science, East China Normal University, Shanghai, China
[3]Nuclear Medicine Department, Southwest Hospital (the First Affiliated Hospital), Third Military Medical University (Army Medical University), Chongqing, China
[4]RadioDynamic Medical, Shanghai, China

## Abstract:

Background: Diagnostic I-131 whole-body planar imaging in differentiated thyroid cancer (DTC) is inherently count-limited due to low administered activity, and image quality deteriorates further under rapid acquisition protocols. Conventional deep learning approaches trained on count-limited diagnostic data struggle to recover true signal from noise, as the training targets themselves lack sufficient photon statistics.

Purpose: To investigate whether high-quality post-therapeutic (PT) I-131 scans can serve as superior training references for enhancing extremely sparse diagnostic planar images through transfer learning and input-image addition strategies.

Methods: In this single-center prospective study, 111 patients were enrolled and classified into PT training (n = 19), Diagnostic follow-up (FU) training (n = 58), and FU paired test (n = 34) cohorts. Five Restormer-based Transformer models were trained under distinct strategies: FU_DL×2 using FU data as label; PT_DL×2 transfer-learned on PT data; PT_DL×1+ and PT_DL×2+ applying input-image addition; and PT_DL×6 as a PT-only reference. Test-time processing employed a two-step de-normalization and Poisson re-scaling procedure to preserve noise statistics during cross-domain intensity matching. Image quality was assessed using PSNR, SSIM, and a blinded dual-reader subjective evaluation on all 34 test cohort.

Results: Among all objective methods, PT_DL×2 achieved the highest PSNR (34.4±3.6 dB) and SSIM (0.81±0.07), while PT_DL×1+ and PT_DL×2+ yielded comparable metrics (33.8 and 33.3 dB; both SSIM 0.80). . Notably, FU_DL×2 ranked below the unprocessed fast scan in both PSNR (30.3 vs. 32.0 dB) and SSIM (0.76 vs. 0.79). Subjectively, PT_DL×2+ received the highest ratings for overall image quality (4.54±0.60) and diagnostic confidence (4.54±0.60), surpassing the reference standard (4.19±0.83 and 4.31±0.81, respectively; P = 0.005 and P = 0.008). FU_DL×2 and PT_DL×6 scored lowest among all methods.

Conclusion: High-quality post-therapeutic I-131 scans provide effective training priors for reconstructing extremely sparse images. Input-image addition and cross-domain transfer learning from count-rich PT data outperform direct training on count-limited diagnostic data. The proposed approach supported 2× scan acceleration and yielded objective and subjective image-quality measures that matched or exceeded those of the unaccelerated standard scan for selected PT-derived configurations.

## Introduction

Differentiated thyroid cancer (DTC) is the most prevalent endocrine malignancy, with a rapidly rising global incidence over the past three decades [1]. Iodine-131 (I-131) whole-body planar imaging plays an indispensable role in disease staging, detection of locoregional and distant metastases, and evaluation of therapeutic response following thyroidectomy [2,3]. Despite its critical diagnostic utility, the technique is fundamentally constrained by the trade-off between image quality and patient comfort. Diagnostic follow-up (FU) scans, typically performed with low administered activities (e.g., 185 MBq), are inherently count-limited, resulting in poor signal-to-noise ratio (SNR). Furthermore, the clinical imperative to improve workflow efficiency and patient compliance necessitates rapid acquisition protocols. However, accelerated acquisition drastically reduces photon counts, rendering data extremely sparse and often compromising the detectability of subtle lesions and anatomical landmarks.

Deep learning (DL) has emerged as a promising solution for enhancing low-count nuclear medicine images. Advanced architectures, including convolutional neural networks (CNNs) and Transformer-based models, have demonstrated remarkable efficacy for denoising low-count SPECT, enabling accelerated bone scintigraphy, and improving low-dose PET image fidelity. The prevailing paradigm in this domain involves training models on paired datasets comprising high-count reference images and computationally down-sampled low-count counterparts, typically generated through Poisson resampling. This approach seeks to learn a robust mapping from noisy to clean images, theoretically restoring diagnostic quality. However, a critical and underexplored limitation of this methodology is its reliance on the assumption that the reference images possess sufficient photon statistics to define anatomical structures reliably. When the standard-acquisition reference images themselves are count-starved, as is frequently the case in diagnostic I-131 imaging, the training targets lack the necessary structural fidelity to guide effective reconstruction.

This study hypothesizes that high-quality post-therapeutic (PT) I-131 scans—acquired with substantially higher count rates—can serve as superior training references for reconstructing extremely sparse diagnostic images. We introduce two methodological innovations to exploit this cross-domain knowledge: (1) a cross-domain transfer learning strategy that fine-tunes models pre-trained on diagnostic data using PT scans, thereby leveraging rich anatomical context; and (2) an input-image addition strategy that preserves authentic photon-count signals while the model recovers missing structural details. These approaches address the fundamental limitation of training on count-starved data and enable clinically viable 2× scan acceleration without compromising diagnostic quality.
The primary contributions of this work are threefold: (1) empirical demonstration that PT-derived models with input-image addition outperform conventional same-domain training in both objective and subjective image quality; (2) a systematic framework for cross-domain knowledge transfer in nuclear medicine imaging; and (3) evaluation of 2× accelerated acquisition using objective image-quality metrics and subjective reader scores, with potential implications for patient comfort and workflow efficiency.

## Material and methods

### 2.1 Patient Population

This single-center, prospective study was conducted in accordance with the Declaration of Helsinki and received approval from the Institutional Review Board of Southwest Hospital. All participating patients provided written informed consent. Patients aged ≥18 years with histopathologically confirmed DTC scheduled for I-131 whole-body planar imaging were eligible for enrollment. Exclusion criteria included pregnancy, concurrent non-thyroidal malignancy with systemic metastasis, prior cervical external-beam radiotherapy, and incomplete imaging data.
Between November 2024 and December 2025, 111 patients were prospectively enrolled and assigned to three cohorts based on their total photon counts and acquisition protocol (Table 1). The PT Training cohort comprised 19 patients whose single-scan total counts were consistent with post-therapeutic I-131 imaging performed 3–7 days after administration of 3.7–7.4 GBq. The FU Training cohort comprised 58 patients whose total counts were consistent with diagnostic I-131 imaging at 185 MBq administered activity. All training scans were acquired at a standard bed speed of 15 cm/min. The Paired Test cohort comprised 34 patients distinct from both training cohorts, each of whom underwent two consecutive whole-body scans without repositioning: a reference scan at 15 cm/min immediately followed by a rapid scan at 30 cm/min (2× acceleration). The paired scans were acquired consecutively without repositioning using a fixed standard-then-rapid order to enable within-subject comparison and pixel-wise spatial correspondence without deformable registration; inter-scan timing and any decay effects were not separately quantified.  were included in both quantitative and subjective evaluations.

Table 1. Patient demographics

| | PT Training (n=19) | FU Training (n=58) | Paired Test (n=34) |
|---|---|---|---|
| Age (years) | 42.7 ± 12.1 (26–65) | 41.7 ± 14.8 (18–73) | 40.6 ± 13.4 (19–70) |
| Sex (M/F) | 6/13 | 18/40 | 13/21 |
| Administered activity | 3.7–7.4 GBq | 185 MBq | 185 MBq |

## 2.2 Image Acquisition

All imaging was performed on a Siemens Symbia T6 SPECT/CT system (Siemens Healthineers, Erlangen, Germany) equipped with a high-energy (HE) collimator. The energy window was centered at 364 keV with a width of ±15%. Whole-body planar images were acquired in a 256 × 1024 matrix extending from the vertex to mid-thigh. For the training cohorts (PT and FU), all scans were acquired at the standard bed speed of 15 cm/min, corresponding to a scan duration of approximately 20–30 minutes. For the Paired Test cohort, the standard 15 cm/min reference scan was followed immediately by a 30 cm/min rapid scan lasting approximately 10–15 minutes, with the patient remaining in the same position throughout.

## 2.3 Training Data Preparation

### 2.3.1 Poisson Down-Sampling

Paired training data were generated by computationally thinning standard-acquisition images through independent Poisson resampling [19]. For a high-count reference image  , a simulated low-count image was produced as:

where α is the down-sampling factor. For models trained on FU dataset, α = 2 was applied to match the test-cohort rapid acquisition speed of 30 cm/min. For models trained on PT dataset, α = 6 was applied to exploit the superior photon statistics of PT scans while simulating extreme sparsity. This formulation faithfully replicates the Poisson counting statistics inherent to nuclear medicine detection [20]. Each scan was independently resampled five times with distinct noise realizations, yielding 290 paired training samples from the FU cohort (58 × 5) and 95 paired samples from the PT cohort (19 × 5). For the transfer learning configuration, the 95 PT-derived paired samples were used to fine-tune a model pre-trained on FU data.

### 2.3.2 Image Normalization and De-normalization

During training, input and target images are independently normalized to [0, 1] using their own per-image maximum and minimum values. This independent normalization strategy decouples the denoising objective from absolute intensity variations, enabling the model to learn relative count distributions independent of administered activity levels.

At inference, the model output in [0, 1] undergoes a two-step rescaling to reach the target intensity scale. In the first step, the output is de-normalized to the training-domain intensity range. Models trained with Poisson down-sampling factor $\alpha = 2$ de-normalize to 2× the input intensity range; models trained with $\alpha = 6$ de-normalize to 6×. In the second step, the de-normalized image undergoes Poisson down-sampling with retention probability $p = n / \alpha$ to reach the target ×n scale, preserving Poisson noise statistics at the target intensity level. This two-step design replaces linear amplitude scaling with a physically grounded Poisson thinning operation that respects the statistical nature of photon-count-limited data.

## 2.4 Network Architecture

We employed the Restormer architecture [18], a Transformer-based encoder-decoder network that achieves state-of-the-art performance on image restoration tasks while maintaining computational efficiency through its Multi-Dconv Head Transposed Attention (MDTA) mechanism. Unlike standard Vision Transformers, whose self-attention exhibits quadratic complexity with respect to spatial resolution, Restormer performs attention across channel dimensions, yielding linear complexity and making it well-suited for high-resolution medical images.

The network follows a hierarchical U-shaped encoder-decoder design with skip connections (Figure 1). The input is a single-channel noisy planar image, and the output is a learned residual correction added to the input. The encoder comprises four stages with progressively increasing channel dimensions (48 → 96 → 192 → 384), corresponding Transformer blocks (4, 6, 6, 8), and attention heads (1, 2, 4, 8). Downsampling between stages is achieved via PixelUnshuffle operations, which reduce spatial resolution by a factor of 2 while doubling channel count, preserving feature capacity without information loss. The decoder employs three stages of PixelShuffle upsampling to reconstruct spatial resolution, with skip connections concatenating encoder features to decoder features at corresponding resolutions, followed by 1 × 1 convolution for channel compression. A final refinement stage with four Transformer blocks processes the aggregated features, and a 1 × 1 convolution projects the output to a single-channel residual map.

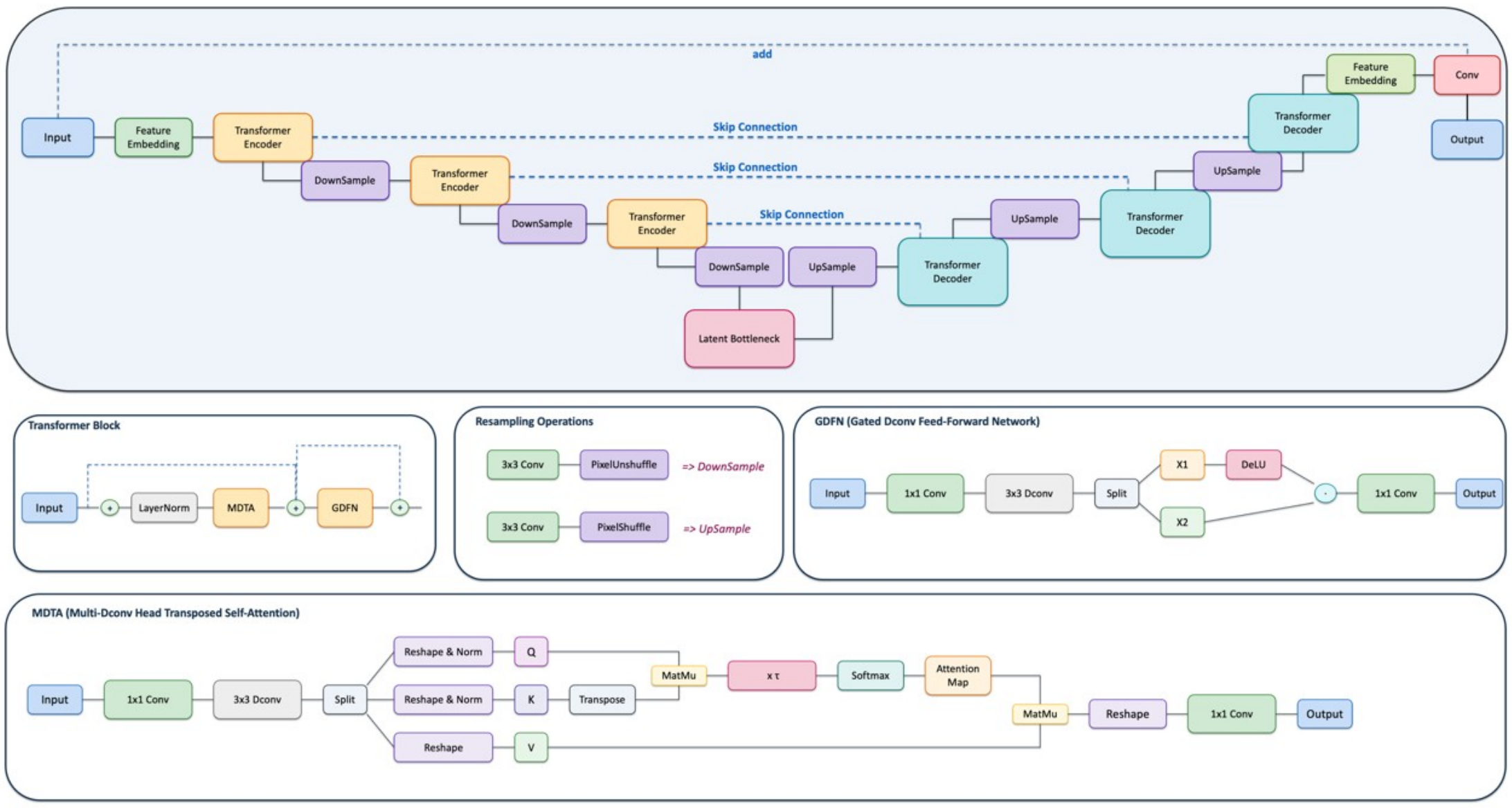

**Fig. 1** Restormer architecture for I-131 planar image reconstruction. The network employs a U-shaped encoder-decoder design with skip connections. MDTA: Multi-Dconv Head Transposed Attention; GDFN: Gated-Dconv Feed-Forward Network. Downsampling via PixelUnshuffle; upsampling via PixelShuffle.

The MDTA module, the core attention mechanism of Restormer, computes cross-covariance attention across channel dimensions rather than spatial dimensions. Query, key, and value projections are generated through 1 × 1 convolution followed by 3 × 3 depth-wise convolution, reshaped by attention heads, and multiplied to form a channel-wise attention map encoding global dependencies. A learnable temperature parameter $\tau$ controls attention sharpness, and a final 1 × 1 convolution projects attended features back to the original dimension. This channel-wise attention captures long-range anatomical dependencies—such as symmetric organ uptake patterns and skeletal structure—without the computational burden of spatial self-attention.

Each Transformer block also contains a Gated-Dconv Feed-Forward Network (GDFN), in which features are expanded by a factor of 2.66, processed by 3 × 3 depth-wise convolution, and split into two gating branches combined via GELU-gated element-wise multiplication [21]. Both MDTA and GDFN modules are applied in a pre-norm residual configuration with Layer Normalization.

The network was trained for 3,000 epochs with the Adam optimizer ($\beta_1 = 0.9$, $\beta_2 = 0.999$), a batch size of 4, and random 256 × 256 crops taken from whole-body images. The composite loss function combined L1 pixel-wise error and SSIM-based perceptual loss ( = 1.0,  = 0.1) with an 11 × 11 Gaussian window [22].

The initial learning rate of $4 \times 10^{-4}$ was halved every 800 epochs. For the transfer learning configuration (described in Section 2.5), a model pre-trained on FU data was fine-tuned on PT data for an additional 1,000 epochs with a reduced learning rate of $1 \times 10^{-4}$. All experiments were conducted on two NVIDIA RTX3090 GPUs.

## 2.5 Model Design and Nomenclature

### 2.5.1 Naming Convention

Each model is named according to a convention that encodes its training data source, output intensity scale, and inference-time processing. The prefix (FU_DL or PT_DL) identifies whether the model was trained using diagnostic follow-up scans or post-therapeutic scans as the label. The suffix ×n specifies the final output intensity scale relative to the input image, implemented through the two-step de-normalization and Poisson re-scaling procedure described in Section 2.3 ($\alpha = 2$ for FU_DL; $\alpha = 6$ for PT_DL). Models bearing the + suffix apply input-image addition at inference: the final output is formed as the element-wise sum of the model's processed prediction and the original input image, preserving the authentic photon-count distribution of the acquisition while the model contributes recovered structural detail.

### 2.5.2 Model Configurations

Five model configurations were designed to test the hypothesis that PT-derived training—with or without input-image addition—can outperform conventional same-domain FU training when applied to extremely sparse diagnostic images.

The baseline model, FU_DL×2, represents the conventional same-domain paradigm: it is trained and evaluated on data from the same distribution. This model uses FU scans as training labels with 2× Poisson down-sampling, matching the test-cohort rapid acquisition speed, and applies test-time de-normalization at 2× the input intensity range.

PT_DL×2, PT_DL×1+, and PT_DL×2+ share a common PT-trained backbone, derived through transfer learning to address the limited size of the PT cohort (n = 19). This backbone is initialized from fully trained FU_DL×2 weights and fine-tuned on PT scans with $\alpha = 6$ Poisson down-sampling, then used to generate three inference variants that differ only in post-processing. PT_DL×2 applies ×2 de-normalization with direct output. PT_DL×1+ applies ×1 de-normalization with input-image addition. PT_DL×2+ applies ×2 de-normalization with input-image addition. None of the variants requires additional training; all three share the same backbone weights and differ only in the de-normalization factor and whether the addition step is applied.

PT_DL×6 is trained directly on PT scans with 6× Poisson down-sampling and ×6 de-normalization. It provides a reference for PT-only training under small-sample conditions and contextualizes the contributions of the transfer learning and addition strategies.

## 2.6 Evaluation

Seven image series were generated per test patient for within-subject comparison: PT_DL×2+, PT_DL×1+, PT_DL×2, FU_DL×2, PT_DL×6, the fast scan, and the standard scan. Image fidelity was assessed using Peak Signal-to-Noise Ratio (PSNR) and Structural Similarity Index Measure (SSIM) [22], with the standard scan serving as reference. Both metrics were computed over the full image. For the subjective reader study, two board-certified nuclear medicine physicians with at least seven years of I-131-specific experience independently evaluated all 238 anonymized image series in independently randomized order. Readers were blinded to all clinical information, acquisition parameters, and reconstruction method, and could freely adjust window settings on calibrated diagnostic monitors. Each image was rated on a 5-point Likert scale across two dimensions: overall image quality and diagnostic confidence. Scale anchors ranged from 1 to 5, with 1 indicating non-diagnostic quality/anatomy indistinguishable and 5 indicating excellent delineation ideal for clinical interpretation.

## 2.7 Statistical Analysis

All analyses were performed with $\alpha = 0.05$ (two-sided). Normality of Likert score distributions was assessed using the Shapiro-Wilk test. Comparisons among the seven image series were performed using the Friedman rank test, and post-hoc pairwise comparisons used the Wilcoxon signed-rank test; effect sizes (r) were planned with conventional thresholds (<0.1 negligible, ≥0.1 small, ≥0.3 medium, ≥0.5 large) and should be reported alongside the corresponding test statistics where calculated. Inter-rater agreement was assessed using quadratically weighted Cohen's kappa (κ) with 95% confidence intervals. Agreement was interpreted using the Landis and Koch thresholds (κ > 0.80 very good, 0.61–0.80 good, 0.41–0.60 moderate, 0.21–0.40 fair, and ≤ 0.20 poor) [23]. Paired Wilcoxon tests compared raw versus reconstructed PSNR and SSIM values.

# 3. Results

## 3.1 Quantitative Assessment

PSNR and SSIM were computed against the reference standard scan for all 34 patients in the test cohort (Table 2). PT_DL×2 achieved the highest PSNR (34.4 ± 3.6 dB) and SSIM (0.81 ± 0.07) among all methods. PT_DL×1+ and PT_DL×2+ followed closely (33.8 and 33.3 dB, respectively; both SSIM 0.80). All three PT-derived models significantly outperformed the fast scan in PSNR ($P < 0.001$).

FU_DL×2—trained on count-limited FU data—ranked below the unprocessed fast scan in PSNR (30.3 vs. 32.0 dB; SSIM: 0.76 vs. 0.79), indicating that same-domain training on sparse data degraded image quality relative to the raw acquisition. PT_DL×6 showed the poorest quantitative performance among all methods (PSNR: 23.0 ± 3.8 dB; SSIM: 0.68 ± 0.06), consistent with its output retaining the full 6× training-scale intensity without Poisson re-scaling to the diagnostic range.

Table 2. Quantitative Metrics Across Image Series (n = 34)

| | PSNR | SSIM |
|---|---|---|
| PT_DL×1+ | 33.8±3.6 | 0.80±0.07 |
| PT_DL×2 | 34.4±3.6 | 0.81±0.07 |
| PT_DL×6 | 23.0±3.8 | 0.68±0.06 |
| FU_DL×2 | 30.3±3.7 | 0.76±0.07 |
| PT_DL×2+ | 33.3±3.6 | 0.80±0.06 |
| Fast scan | 32.0±3.5 | 0.79±0.07 |

## 3.2 Subjective Radiologist Assessment

Inter-rater agreement ranged from good to very good for all algorithms (quadratically weighted Cohen's κ: 0.674-0.933). Subjective assessment results are summarized in Table 3.
PT_DL×2+ achieved the highest scores across two dimensions, exceeding the reference standard by 0.35 points in overall image quality (4.54 ± 0.60 vs. 4.19 ± 0.83; P = 0.005) and by 0.23 points in diagnostic confidence (4.54 ± 0.60 vs. 4.31 ± 0.81; P = 0.008). PT_DL×1+ had the second-highest mean scores in both metrics (overall quality: 4.28 ± 0.66; diagnostic confidence: 4.26 ± 0.65). Relative to the reference standard, overall image quality was numerically higher (4.28 ± 0.66 vs. 4.19 ± 0.83), whereas diagnostic confidence was slightly lower (4.26 ± 0.65 vs. 4.31 ± 0.81). PT_DL×2 (overall quality: 4.18 ± 0.59; diagnostic confidence: 4.19 ± 0.69) performed comparably to the reference standard.
Notably, FU_DL×2 ranked second-to-last among all methods, falling below the unprocessed fast scan, consistent with its poor quantitative performance (PSNR: 30.3 dB, the lowest among all DL methods excluding PT_DL×6; Table 2). PT_DL×6, the PT-only trained model without transfer learning or addition, performed poorest among all methods (overall quality: 2.25 ± 0.65; diagnostic confidence: 1.99 ± 0.68), reflecting the inadequacy of the small PT cohort (n = 19) for direct training without transfer learning and the adverse effect of applying ×6 intensity scaling to diagnostic-range inputs.

Table 3. Subjective Assessment Across Two Dimensions (n = 34 patients, 2 raters)

| | Overall Image Quality | Diagnostic Confidence |
|---|---:|---:|
| PT_DL×2+ | 4.54 ± 0.60* | 4.54 ± 0.60* |
| PT_DL×1+ | 4.28 ± 0.66 | 4.26 ± 0.65 |
| Standard scan | 4.19 ± 0.83 | 4.31 ± 0.81 |
| PT_DL×2 | 4.18 ± 0.59 | 4.19 ± 0.69 |
| Fast scan | 2.98 ± 1.24* | 2.97 ± 1.39* |
| FU_DL×2 | 2.72 ± 1.27* | 2.82 ± 1.26* |
| PT_DL×6 | 2.25 ± 0.65* | 1.99 ± 0.68* |

All data are expressed as the mean ± SD. *: indicates $p < 0.05$, and Standard scan serves as a quality reference

Figure 2 presents a representative case from the test cohort, showing all seven image series for visual comparison. The fast scan exhibits pronounced noise and poorly defined anatomical boundaries, particularly in regions of low background uptake. FU_DL×2 introduces structured artifacts and false texture in signal-void regions, consistent with its poor quantitative and subjective scores. PT_DL×2, PT_DL×1+, and PT_DL×2+ demonstrate progressively stronger structural recovery, with PT_DL×2+ most closely approximating the reference standard in both lesion clarity and background homogeneity.

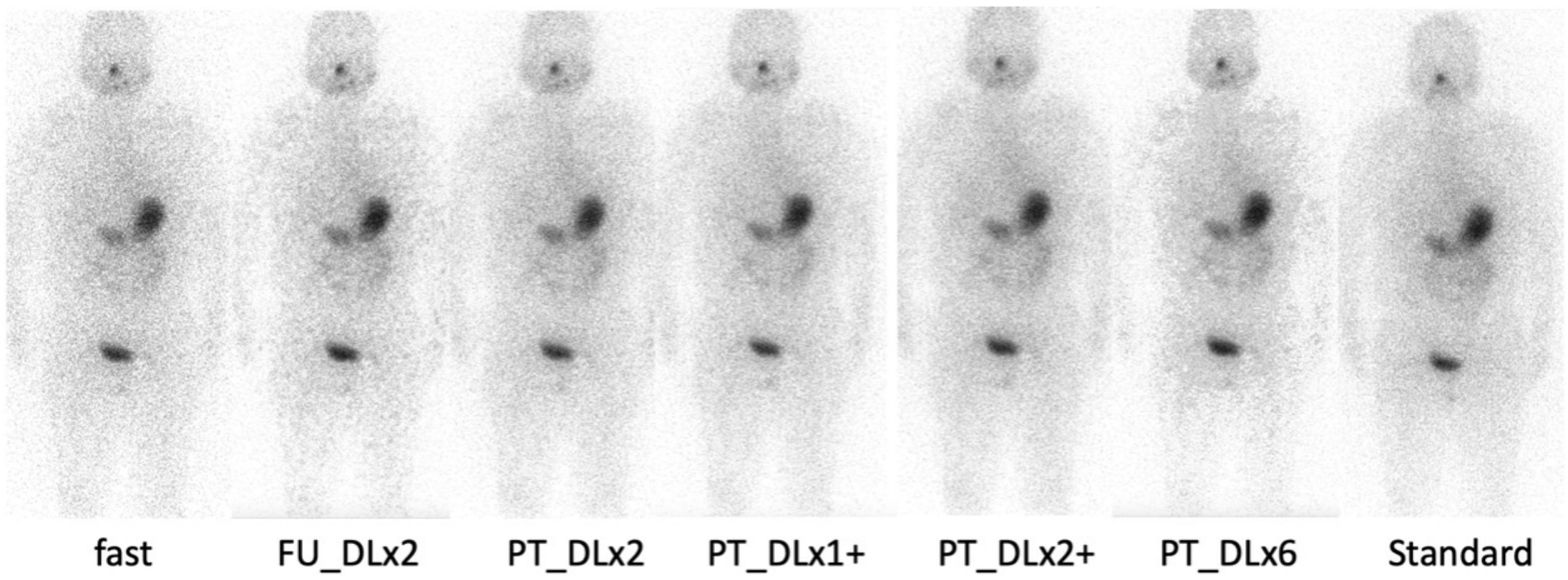

**Fig. 2** Representative case from the test cohort. Images shown from left to right: fast scan (30 cm/min), FU_DL×2, PT_DL×2, PT_DL×1+, PT_DL×2+, PT_DL×6, and the reference standard scan (15 cm/min). All images are displayed with identical window settings.

## 4. Discussion

This study demonstrates that high-quality post-therapeutic I-131 scans can serve as effective training references for reconstructing extremely sparse diagnostic planar images, and that input-image addition is a critical safeguard against noise amplification during cross-domain inference. FU_DL×2, trained on count-limited FU data following the conventional same-domain paradigm, not only failed to improve image quality but degraded it relative to the unprocessed fast scan (PSNR: 30.3 vs. 32.0 dB; subjective quality: 2.72 vs. 2.98). In contrast, PT-derived models with input-image addition produced images that matched or exceeded the reference standard. PT_DL×2+ achieved the highest subjective ratings (overall quality: 4.54 ± 0.60 vs. 4.19 ± 0.83 for the reference standard; (overall quality: 4.28 ± 0.66 vs. 2.72 ± 1.27; diagnostic confidence: 4.26 ± 0.65 vs. 2.82 ± 1.26; both P < 0.001). PT_DL×2, despite achieving the highest PSNR (34.4 dB), produced subjectively inferior images (overall quality: 4.18 ± 0.59; diagnostic confidence: 4.19 ± 0.69) compared to both addition models (PT_DL×1+: 4.28 and 4.26; PT_DL×2+: 4.54 and 4.54). Together, PT-based training labels, input-image addition, and Poisson-grounded intensity re-scaling form an effective framework for extremely sparse data reconstruction.

The degradation produced by FU_DL×2 likely reflects a specific interaction between the biodistribution of I-131 and the learning dynamics of deep image restoration models. Unlike bone-seeking tracers such as Tc-99m MDP, which produce a continuous skeletal outline even at reduced count densities, I-131 accumulates selectively in thyroid remnants and iodine-avid metastases, with minimal uptake in background soft tissues. At standard acquisition speeds (15 cm/min), sufficient counts are accumulated to define gross anatomical context. When acquisition is accelerated to 30 cm/min, however, photon counts are halved, and large regions of the image field register single-digit counts or zero. In these signal-void regions, the training targets are dominated by Poisson noise, and the model cannot distinguish between the stochastic absence of counts and genuine anatomical boundaries. It learns to interpolate structured noise patterns that readers perceive as artifacts, whereas the fast scan simply displays the underlying Poisson uncertainty. This interpretation is consistent with observer studies in cardiac SPECT, where DL-based denoising at very low counts has been shown to degrade defect detection despite improved noise metrics [8,9].

PT-trained models avoid this problem because their training targets were derived from higher-activity post-therapeutic acquisitions and, for training-data preparation, used a Poisson down-sampling factor of $\alpha = 6$ to simulate extreme sparsity, providing structurally richer reference images in which full-body morphological context is more clearly defined. When applied to sparse diagnostic data, these models contribute learned structural context that constrains the interpretation of sparse focal signals, analogous to the role of anatomical priors in Bayesian image reconstruction. The input-image addition mechanism provides a further safeguard: by summing the model output with the original input, the authentic photon-count distribution of the acquisition is preserved, and the model is restricted to contributing only missing structural context rather than generating the image directly. This likely explains why PT_DL×2, despite achieving the highest PSNR, produced subjectively inferior images compared to both addition models (PT_DL×1+ and PT_DL×2+). When a model generates every pixel from learned features, even small deviations from the true count distribution accumulate into perceptible textural inconsistencies. Anchoring the output to the original measurement eliminates this source of degradation.

The finding that DL-based denoising can degrade diagnostic quality when training targets are themselves count-limited is consistent with observations in other modalities. In low-dose CT, models trained on noisy reference data have been shown to produce plausible image textures while reducing lesion conspicuity [25]. In cardiac SPECT, Sohlberg et al. reported that post-reconstruction DL denoising at very low count densities degraded myocardial perfusion defect detection despite reducing image noise [8,9]. The present study extends this evidence to planar I-131 imaging and identifies a specific physical mechanism: the absence of structural signal in count-starved background regions, which is a direct consequence of the highly selective uptake pattern of radioiodine.

The cross-domain strategy aligns with broader developments in medical image reconstruction. Zhou et al. showed that federated transfer learning could adapt PET denoising models across institutions with heterogeneous dose protocols [14], and Huang et al. demonstrated that frequency-domain alignment could bridge domain gaps in CT denoising [27]. The present work extends this paradigm to a more extreme post-therapeutic to diagnostic I-131 imaging, encompassing a large difference in administered activity and a Poisson simulation factor of $\alpha = 6$ rather than a directly measured six-fold count-rate ratio. The introduced input-image addition works as a simple, architecture-agnostic strategy for preserving measurement fidelity during cross-domain inference.

Among studies specifically addressing accelerated planar scintigraphy, Pan et al. and Qi et al. demonstrated that content-attention networks could enhance bone scans acquired at 2–3× acceleration to match standard-acquisition quality [11–13]. The present work differs in two respects: the tracer (I-131 vs. Tc-99m MDP), which fundamentally alters the noise structure as discussed above, and the training strategy (cross-domain PT labels vs. same-domain standard labels). The observation that PT-derived training was effective despite the domain mismatch suggests that count statistics, rather than domain alignment per se, may be the dominant determinant of training data quality for low-count image restoration. Whether this finding generalizes to other tracers and acquisition protocols remains to be investigated.

This study has several limitations. The single-center design and modest sample size (n = 111; reader study n = 34) may limit generalizability, although the Friedman test confirmed significant overall differences among algorithms ($P < 0.001$). The reader study employed Likert-scale assessments rather than task-based metrics such as lesion detection accuracy via localization receiver operating characteristic analysis; the latter would provide more direct evidence of clinical non-inferiority. The limited size of the PT cohort (n = 19) constrained direct PT-only training, and both PT_DL×2 and PT_DL×6 should be regarded as pragmatic compromises rather than fully optimized strategies. The optimal Poisson re-scaling factor was not exhaustively characterized, and adaptive per-image strategies may extend the effective range beyond ×2. Computational deployment considerations, including inference latency and PACS integration, were not evaluated.

## 5. Conclusion

Training on count-limited FU data substantially degraded image quality relative to the unprocessed fast scan. In this single-center test cohort, PT-derived models with input-image addition reversed that degradation; PT_DL×2+ exceeded the reference standard on both subjective dimensions, supporting further evaluation of 2× accelerated acquisition. These findings establish that high-quality PT scans, combined with input-image addition and Poisson re-scaling, provide an effective framework for extremely sparse planar image reconstruction in DTC surveillance.